\documentclass[letterpaper,twocolumn,12pt]{article}
\usepackage[utf8]{inputenc}
\usepackage[english]{babel}
\usepackage{graphicx}
\usepackage{epsf}
\usepackage{rotating}
\usepackage{afterpage}
\usepackage{scs}
\usepackage{amsmath,amsfonts,amssymb,amsthm}
\usepackage{epsfig,epstopdf}
\usepackage{hyperref}
\usepackage{tkz-graph}

\usetikzlibrary{arrows}

\begin{document}

\newtheorem{defn}{Definition}

\title{A Markov Model for Ontology Alignment}

\author{Michael E. Cotterell and Terrance Medina \\
Department of Computer Science \\
The University of Georgia \\
Athens, GA 30602-7404 \\
mepcott@uga.edu, medinat@uga.edu
}

\maketitle

\begin{abstract}
The explosion of available data along with the need to integrate and utilize that
data has led to a pressing interest in data integration techniques. In terms of
Semantic Web technologies, Ontology Alignment is a key step in the process of 
integrating heterogeneous knowledge bases. In this paper, we present
the Edge Confidence technique, a modification and improvement over the popular
Similarity Flooding technique for Ontology Alignment.

\end{abstract}

\section{Introduction}


As database technologies become increasingly diverse, the need to integrate those technologies has become ever more important. 
The heterogeneous data problem describes a common situation in which multiple data sources with incompatible descriptions and data types must be integrated for use by a single application. 
This problem is often encountered in the Semantic Web, which attempts to view the entire Internet as a unified database. 
A fundamental step in solving the heterogeneous data problem is the production of an \textit{alignment} that can tell the client how to equate the descriptions of two heterogeneous data sources.

An ontology alignment may be informally described as a set of correspondences between semantically related terms in two heterogeneous input ontologies. 
Each correspondence is qualified with a confidence level, $[0,1]$.
Alignment is useful in data integration tasks dealing with what is sometimes referred to as the semantic heterogeneity problem. 
It helps in the automation of various important tasks, most important of which is schema merging, enabling the knowledge and data expressed in the input ontologies to inter-operate.

Ontology alignment algorithms are typically aggregations of multiple basic techniques, which may be classified as lexical, semantic, structural etc~\cite{euzenat:2007:ontology}.
In this paper, we consider the structural technique of Similarity Flooding and present an improvement to that technique. \\

\noindent The rest of this paper is organized as follows: 
the remainder of this section will cover some of the background material related to the Semantic Web and ontologies, the cases for fully automated alignment, and some of the background material on the statistical methods used in this paper; 
Section~\ref{sec:approach} outlines the approach and implementation details; 
the evaluation and benchmarks are explained in Section~\ref{sec:eval}; 
results are presented in Section~\ref{sec:results}; and, 
conclusions and possible future work are outlined in Section~\ref{sec:conclusions}.

\subsection{Semantic Web and Ontologies}
\label{subsec:semanticweb}
First, we define some key terms around the Semantic Web and Ontology Alignment.

A \textit{knowledge base} is a kind of database that is designed for decision support systems and expert systems. 
It specifically allows machines to perform deductive reasoning over its elements.
For example, given the instance data ``John is a farmer'' and ``Farmers wear overalls'', a knowledge base reasoner could deduce the new information that ``John wears overalls''.

Knowledge bases (KBs) consist of \textit{entities} and \textit{relations} between those entities. 
They are often represented as sets of triples, that is, a set of two entities and a relationship between those entities. 
For example, the triple ${\{teachers, lesson\_plans, write\}}$ defines the relationship that ``teachers write lesson\_plans''.

\textit{Ontologies} describe the relationships between elements in a knowledge base.
Similar to the notion of schemas in relational databases, ontologies specify the structural relationships amongst the entities and classes of the ontology.

Frequently, queries must be performed across multiple knowledge bases.
This is the case when departments in a large enterprise maintain their own knowledge bases, but must also share information with other departments. 
This also happens in scientific computing, especially in the more domain-specific Informatics areas such as Bioinformatics and Energy Informatics~\cite{cotterell:2012:oei}.
It is also a fundamental requirement of the Semantic Web; users perform queries against the web, and those queries must be performed against knowledge bases that may belong to entities on opposite sides of the globe. 
This leads us to the problem of heterogeneous data. 
When knowledge bases are maintained by different organizations, the ontologies used to describe them will usually be different. 
For instance, one ontology may keep track of `cars' while another keeps track of `automobiles' and still another keeps track of `autoProducts'. 
In fact, the list could go on and on.

\textit{Ontology Alignment} is the process of equating two heterogeneous ontologies by finding valid correspondences between their sets of elements. 
For instance, an alignment between an auto parts manufacturer and an auto dealership might tell us with 85\% certainty that ``anti\_lock\_brakes'' and ``ABS'' represent the same thing.

This is important because it allows a query management system to translate the terminology of a user's query into the terminology used by many different ontologies and thereby to query heterogeneous knowledge bases.

\subsection{Why do we need fully automated alignment?}
\label{subsec:automated}
Fully-automated alignment techniques (i.e., ontology alignment performed without any input or approval from a human user) represent the ideal scenario for many of the use cases already discussed. 
For example, in a Semantic Web query, the details of equating ontology elements across heterogeneous KBs should remain completely invisible to the user, who simply wants to know the show-times of movies in his or her zip code, which star actors with a maximum Kevin-Bacon-Distance of three~\cite{hayes:2000:graph}.
On the other hand, in an Enterprise Integration context much of the information to be integrated is extremely domain specific and even organization specific, and possibly only known by a handful of organization veterans. 
In such a case, a fully-automated alignment process may not be entirely feasible.

In any case, fully-automated alignment is an as-yet-unattained goal. 
In this paper we present an improvement to Similarity Flooding, a structural alignment technique. 
However, both the original algorithm and our improvement rely on comparison against an ideal alignment, produced by a human, for evaluation of the quality of the alignment. 
For our evaluation, we have used the Semantic Evaluation At Large Scale (SEALS) automated testing platform, which subjects an algorithm to a battery of alignment test, and produces \textit{precision} and \textit{recall} scores based on a comparison of its results against an ideal alignment result.

\section{Related Work}

Although the problem of fully-automated ontology alignment is far from being solved, there has been much work accomplished around fundamental techniques for element matching and aggregation of those matching techniques~\cite{euzenat:2007:ontology}.
One class of approaches attempts to find element matches based on the relative similarity or dissimilarity of the actual labels given to those elements. 
Some examples of these string-matching techniques are the Jaro-Winkler measure~\cite{winkler:1999:state}, the Levenshtein Distance~\cite{yujian:2007:levenshtein} and Latent Semantic Indexing~\cite{benzecri:1973:lsi}.
While useful in many situations, string-based matching techniques suffer from a common shortcoming; similar real-world objects often have very dissimilar names. 
The words ``car'' and ``automobile'' provide a good example.

Another class of techniques makes use of outside resources as aids in the search for good matches between elements. 
Such outside resources include dictionaries or taxonomies such as the WordNet taxonomy~\cite{miller:1995:wordnet}. 
These classes of techniques attempt to produce correspondences between elements that may likely refer to the same real-world objects, but that have very dissimilar names, such as ``car'' and ``automobile''. 
Some examples of these semantic approaches include Information-theoretic similarity~\cite{lin:1998:information}.
The use of outside taxonomies can greatly improve the quality of matching results, but such outside resources are not always readily available.

Yet a third class of matching techniques attempts to exploit the structure of the ontology itself. 
For example, Similarity Flooding~\cite{melnik:2002:similarity} takes a directed graph representation of an ontology and uses neighbor relations between the elements to find matching correspondences between them. 
The idea is that if two elements in heterogeneous ontologies are very similar, then their neighboring elements should also be very similar.

The Similarity Flooding algorithm operates by producing a new graph that represents relationships amongst the entities in each input ontology. 
The algorithm first takes the cross-product of all nodes in both ontologies, producing a single node in the result graph for each pairing of nodes in the input ontologies. 
Edges in the result graph are produced if and only if the original nodes from the input ontologies both shared an edge. 
This results in a \textit{Pairwise Connectivity Graph}.
Finally, weights are added to the edges such all outbound edges from a given node have equal weight, and sum to one. 
The result is a \textit{Propagation Graph}.

Once the Propagation Graph has been generated, the similarity score of each node is generated as follows: each node is assigned an arbitrary initial similarity
score which is refined through an iterative fix-point computation. 
At each iteration the new similarity is equal to the old similarity plus the weighted sum of the similarity of all its neighbors in the propagation graph. 
This fix-point computation proceeds until the new similarity converges to a fixed value.
%
Similarity flooding has been useful as a foundation in other structure-based matching techniques such as anchor flooding, but suffers from a few limitations. 
First, it requires that the edges of the edge-labeled graph representation have identically named labels. 
In the case that corresponding edge labels do not have exactly the same name, but mean the same thing (for example ``hasA'' and ``has\_a'') that information is completely lost to the flooding algorithm. 


\section{Approach \& Implementation}
\label{sec:approach}

The following sections outline and present the details about the model as well as many of the implementation details.

\subsection{Levenshtein Edit Distance}

The Levenshtein distance is a string similarity metric for measuring the difference between two character sequences. 
The distance between two sequences is equal to the number of single-character operations required to transform one sequence into the other~\cite{yujian:2007:levenshtein}. 
The single-character operations are insertion, deletion, and substitution.

\begin{defn}
Let $x$ and $y$ be terms in a single ontology. The label set $\Lambda \left( x, y \right)$ is the set of all labels between hierarchical properties and object properties where $x$ and $y$ are the domain and range of a property, respectively.
\end{defn}

\begin{figure*}
\centering
\begin{equation*}
\mathcal{L} 
\left( \alpha_i, \beta_j \right) = \left\{
	\begin{array}{ll}
   	 	0 &: i=j=0 \\
		i &: j = 0 \text{ and } i > 0 \\
		j &: i = 0 \text{ and } j > 0 \\
		\min 
			\left\{ 
			\begin{array}{l}
				\mathcal{L} \left( \alpha_{i-1}, \beta_j \right) + 1 \\
          		        \mathcal{L} \left( \alpha_i, \beta_{j-1} \right) + 1 \\
          		        \mathcal{L} \left( \alpha_{i-1}, \beta_{j-1} \right) + [\alpha_i \neq \beta_j]
			\end{array} \right. &: \text{else}
     \end{array}
\right.
\end{equation*}
\caption{Levenshtein Edit Distance}
\end{figure*}

\begin{figure*}
\centering
\begin{equation*}
\sigma
\left( \alpha_i, \beta_j \right) = \left\{
\begin{array}{ll}
  1           &: \mathcal{L} \left( \alpha_i, \beta_j \right) = 0 \\
  \frac{3}{4} &: \mathcal{L} \left( \alpha_i, \beta_j \right) = 1 \\
  \frac{1}{\mathcal{L} \left( \alpha_i, \beta_j \right)} &: \text{otherwise} \\
\end{array}
\right.
\end{equation*}
\caption{Edit Similarity Distance}
\end{figure*}

\subsection{Edge Confidence}
These lexical similarity techniques give us a way of assessing the similarity of two entities regardless of any structural relationships they may have within an ontology. 
Recall that one limitation of the Similarity Flooding technique is the necessity that predicates (edge labels when the ontology is represented as a graph) must have names that correspond exactly. 
But this is an unrealistic requirement. 
It is easy to imagine that two unrelated ontologies might, for instance, use predicates labeled ``appearsIn'' and ``actsIn'' to describe the relationship that a certain actor has been in a movie (possibly with Kevin Bacon).

We propose using lexical similarity to quantify a degree of similarity between predicate levels. 
If the similarity is above some arbitrary threshold, we will consider the two predicates to mean the same thing.
In this way, we can include more edges in the propagation graph, which provides more information about structural relationships to the alignment algorithm.

Next we are faced with the problem of assigning an actual value to the edge similarity. 
In the propagation graph for similarity flooding, each outbound edge from a node is given the same weight as all of the other edges from that node, and those weights all sum to one. 
This allows the graph to be described by a row-stochastic matrix. 
We would like to keep that row-stochastic property for our edge confidence implementation, but we would also like to have different weights on each outbound edge,
proportionate to the degree of similarity (and thus the confidence) shared by the ontology predicates represented by the edge in the propagation graph.

To this end, we start with a dissimilarity metric, such as one provided by the Levenshtein distance algorithm. 
Next we derive a complement for this weight by taking its difference from the sum of all outbound edge weights for a particular node. 
Finally, the edge weight is given as the ratio of that complement, and the sum of all complements for outbound edges of a particular node.

This concept is formalized in the definitions that follow.

\begin{defn}
\label{edgeconfidence}
Let $\gamma$ be some threshold. Edge confidence $\Gamma \left( \alpha, \beta \right)$ is the similarity score between the labels of two edges $\alpha$ and $\beta$ if that similarity is greater than $\gamma$, otherwise $0$. That is,
$$ \Gamma \left( \alpha, \beta \right) = \left\{
   	 \begin{array}{ll}
           \frac{1}{\sigma \left( \alpha, \beta \right)} & : \vert \sigma \left( \alpha, \beta \right) \vert \geq \gamma \\
           0                                             & : \mathrm{otherwise}
     \end{array}
   \right.$$
\end{defn}

\subsection{Unnormalized Pairwise Markov Chain}
\begin{figure*}
\centering
\SetVertexNormal[
	Shape = circle,
    LineWidth = 1pt
]
\SetUpEdge[
	lw = 1pt,
    color = orange,
    labelcolor = white
]
\begin{tikzpicture}
   \Vertex[x=0, y=0]{A}
   \Vertex[x=2, y=2]{B}
   \Vertex[x=4, y=0]{C}
   
   \Vertex[x=6, y=0]{D}
   \Vertex[x=8, y=2]{E}
   \Vertex[x=10, y=0]{F}
   \tikzset{EdgeStyle/.style={->}}
   \Edge[label = $m$](A)(B)
   \Edge[label = $n$](B)(C)
   \Edge[label = $o$](C)(A)
   
   \Edge[label = $m'$](D)(E)
   \Edge[label = $n'$](E)(F)
   \tikzset{EdgeStyle/.append style = {bend left}}
   \Edge[label = $o'$](F)(D)
   \tikzset{EdgeStyle/.append style = {bend left}}
   \Edge[label = $p$](D)(F)  
\end{tikzpicture}
\caption{Example Input Ontologies with Object Properties}
\label{fig:input}
\end{figure*}
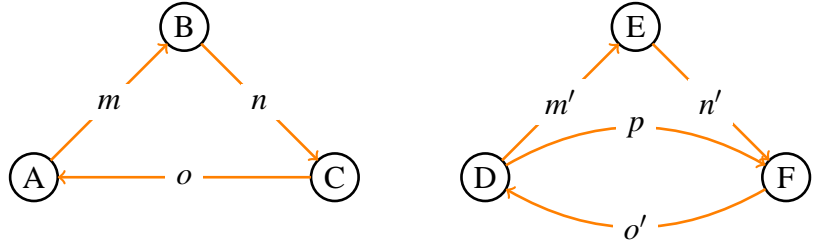

In the definitions that follow, consider the input ontologies presented in Figure~\ref{fig:input}.

A Markov Chain is a mathematical model that represents a system as a set of states and a set of probabilistic transitions between those states. 
Most importantly, Markov processes adhere to the Markov Property, that is, a system's next state depends only on its current state, and not on any of its previous states. 
In that sense, Markov processes are called ``memoryless''~\cite{bcnn:2010:simulation}.

\begin{defn}
An {\bf Unnormalized Pairwise Markov Chain} is a not-necessarily stochastic Markov Chain that satisfies the following property.
For every pairwise grouping of ontological terms between the two input ontologies, there exists a transition in the UPMC
$$ (x, y) \rightarrow (x', y') $$
with probability $\Gamma(\Lambda(x, y), \Lambda(x', y'))$ if and only if
\begin{itemize}
\item there exists an edge from $x$ to $x'$, and
\item there exists an edge from $y$ to $y'$.
\end{itemize}
\end{defn}

The creation of a UPMC creates a directed graph between pairs of concepts that relate pairs based on their structural similarity.
The idea is that a pair of concepts are more likely to be similar if they are structurally similar.
Each pair of ontological terms is connected to other pairs of ontological terms via directed edges that are proportional to similarity of the edges that exist in the input ontologies. 

\subsection{Normalized Pairwise Markov Chain}

In order to take advantage of the convergence properties outlined earlier in this paper, the UPMC needs to have its probability transition matrix converted into a row stochastic form. 
This is done by normalizing the row sums of the matrix such that they sum to one.

\begin{defn}
\label{def:npmc}
a {\bf Normalized Pairwise Markov Chain (NPMC)} is defined as the Markov Chain generated from a UPMC by normalizing the row sums of the matrix such that they sum to one. Let $P$ be the $1$-step probability transition matrix for an UPMC. 
First, determine the sum of the current reciprocal row sums.

$$ M_i = \sum \frac{1}{P_i} $$

\noindent Then, add the reciprocal row sums to each value in the matrix, storing each value in a temporary matrix $T$.

$$ T_{i,j} = -\frac{1}{P_{i,j}} + M_i $$

\noindent Normalize each value by dividing each value in $T$ by its row sum.

$$ P_{i,j} = \frac{T_{i,j}}{\sum T_i} $$
\end{defn}

\noindent The matrix that results from applying the transformation described in Definition~\ref{def:npmc} is \textit{row stochastic} (its rows sum to one).
When viewed as the $1$-step probability transition matrix for a Markov Chain, it is easy to see that the weights on the outgoing edges for each state sum to one.
After the transformation, a NPMC represents a Markov Chain where each pair of ontological terms has a certain probability of being related to other pairs in the chain in proportion to the edge confidences that were calculated earlier.

\subsection{Iterative Approach}

Once approach to finding the stationary distribution of the NPMC is to compute the limiting probabilistic state $\lim_{k \to \infty} \pi P^k$ in an iterative fashion. 
Let $\epsilon$ be some threshold.

$$ A = \pi^t : \pi^t P \pm \epsilon = \pi^{t-1} $$

\noindent The values for the initial similarity distribution $\pi^0$ are taken from the non-structural similarity scores for the ontological terms corresponding to each state.

\subsection{Steady-State Approach}

Another approach to finding the stationary distribution of the NPMC is to compute the limiting probabilistic state $\lim_{k \to \infty} \pi P^k$ directly.
This can be done by solving a left eigenvalue problem: 
$$ \pi = \pi P \Rightarrow \pi (P - I) = 0$$
where the eigenvalue is $1$ and $I$ is the identity matrix. 
Simply solve for $\pi$ by computing the left nullspace of the $P - I$ matrix (appropriately sliced) and then normalizing $\pi$ so that $\vert\vert \pi \vert\vert = 1$.
In order to accomplish this in an easy fashion, the ScalaTion library was used~\cite{miller:2010:scalation}. ScalaTion is an Domain-Specific Emebedded Language (DSEL) for Modeling \& Simulation, written in the Scala programming language~\cite{odersky:2011:spec}.

\subsection{Notes on Convergence}

Since the NPMC model is a MC, certain things can be said about the alignment (stationary distribution), such as~\cite{mitzenmacher:2005:probability}:

\begin{itemize}
\item If the stochastic probability transition matrix P is symmetric then the MC has a unique alignment/stationary distribution.
\item If P is irreducible (but not necessarily aperiodic), then for any 0 < a < 1, the matrix P' = aP + (1 - a)I is stochastic, irreducible and aperiodic, and has the same stationary distribution as P. Note: I is the identity matrix.
Since P' is finite, irreducible, and aperiodic, it's also ergodic and therefore has a unique stationary distribution!
If we can generate such a P' from our NPMC, then we know it has a unique alignment, according to our model.
\end{itemize}

\subsection{Refining Results}

The results generated from using either the iterative approach or the steady-state approach are only somewhat meaningful.
The output of both procedures produces a two-dimensional distribution of similarity scores between the two input ontologies.
Although this satisfies the definition of an alignment as presented earlier in this paper, a user is likely more interested in the set of similarities that yield the best correspondence between the two input ontologies.

Take the alignment distribution generated by either the iterative or steady-state approach and decompose it into an $m \times n$ matrix.
Consider this matrix to be representative of a weighted bipartite graph.
Now, finding the set of similarities that yield the best correspondence between the two ontologies is simply a matter of solving the maximum-weighted bipartite graph matching problem using the generated graph.
In order to accomplish this task, the Hungarian algorithm is used~\cite{kuhn:1955:hungarian}.

%

\section{Evaluation}
\label{sec:eval}

In order to evaluate the model presented in this paper, the bibliographic ontology benchmark provided by 
the Ontology Alignment Evaluation Initiative (OAEI) was used.
This benchmark test library consists of data sets that are built from the bibliographic reference ontology.
Each test includes a reference alignment in order to facilitate the calculation of precision and recall.

In order to evaluate the ontology alignment models presented in this paper, the Semantic Evaluation 
At Large Scale (SEALS) Platform was used~\cite{esteban:2010:executing, wrigley:2010:evaluating}.
The SEALS Platform is an extensible infrastructure that facilitates the remote evaluation of various semantic technologies.
In our evaluation, the SEALS Platform was used to easily compare the models presented in this paper using the OAEI benchmark.
We compared the reference implementation of Similarity Flooding available from Stanford University
to our modified version with the Edge Confidence implementation.

\section{Results}
\label{sec:results}
We compared our Edge Confidence algorithm to the stock Similarity Flooding algorithm using the SEALS Benchmark 1 version 2.0 suite of tests. 
The SEALS platform subjects each algorithm to a battery of alignment tests, then compares the result alignment generated by the algorithm to an ideal alignment, which is included as part of the suite.
For each alignment test, the platform produces scores for \textit{precision} and \textit{recall}.

Precision is defined (Def \ref{def:precision} as the ratio of the number of correct correspondences to the total number of correspondences returned by the algorithm. It should be noted that precision is mainly a penalty against false positives, with no penalty against false negatives. 
For example if there were 100 valid correspondences between two ontologies, a given algorithm could returned only a single correspondence and, if that correspondence were correct, would score a 100\% precision. 

\begin{defn}
\textbf{Precision}
\label{def:precision}
$$
	precision = \frac{|\{valid \} \cap  \{returned \}|}{|\{returned \}|}
$$
\end{defn}

The converse of the precision metric is the recall metric (Def \ref{def:recall}, which provides a penalty for false negatives, but not for false positives. 
Recall considers the ratio of the correct correspondences to the total number of correspondences that should have been returned by the algorithm.

\begin{defn}
\textbf{Recall}
\label{def:recall}
$$
	recall = \frac{|\{valid \} \cap  \{returned \}|}{|\{valid \}|}
$$
\end{defn}

Finally, we consider the F-measure (Def \ref{def:fmeasure}), a metric that provides a balance between precision and recall~\cite{rijsbergen:1979:ir}.. 

\begin{defn}
				\textbf{F-measure}
\label{def:fmeasure}
$$
	F = 2*\frac{precision * recall}{precision + recall}	
$$
\end{defn}

\begin{sidewaysfigure}
\centering
\tiny{\input{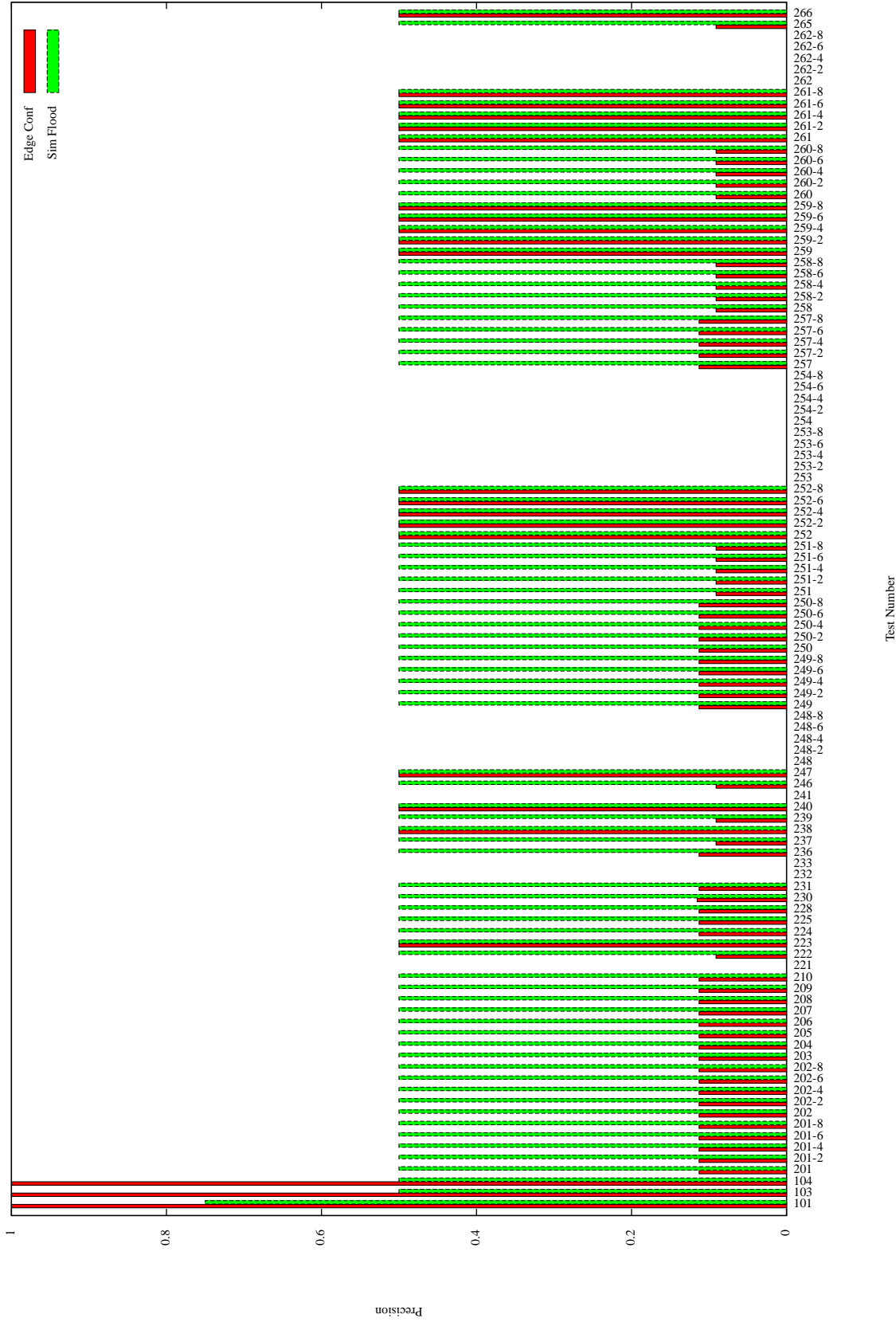}}
\caption{Comparison of Precision}
\label{fig:precision}
\end{sidewaysfigure}

When comparing Similarity Flooding and Edge Confidence on the basis of precision, Similarity Flooding typically outperforms Edge Confidence, as can be seen in Figure~\ref{fig:precision}. 
This reflects the conservative nature of Similarity Flooding.
Because Edge Confidence is more aggressive about including edges, it generates a larger propagation graph, which leads to more correspondences on average, although this will typically include more incorrect results, thus lowering the precision score.

\begin{sidewaysfigure}
\centering
\tiny{\input{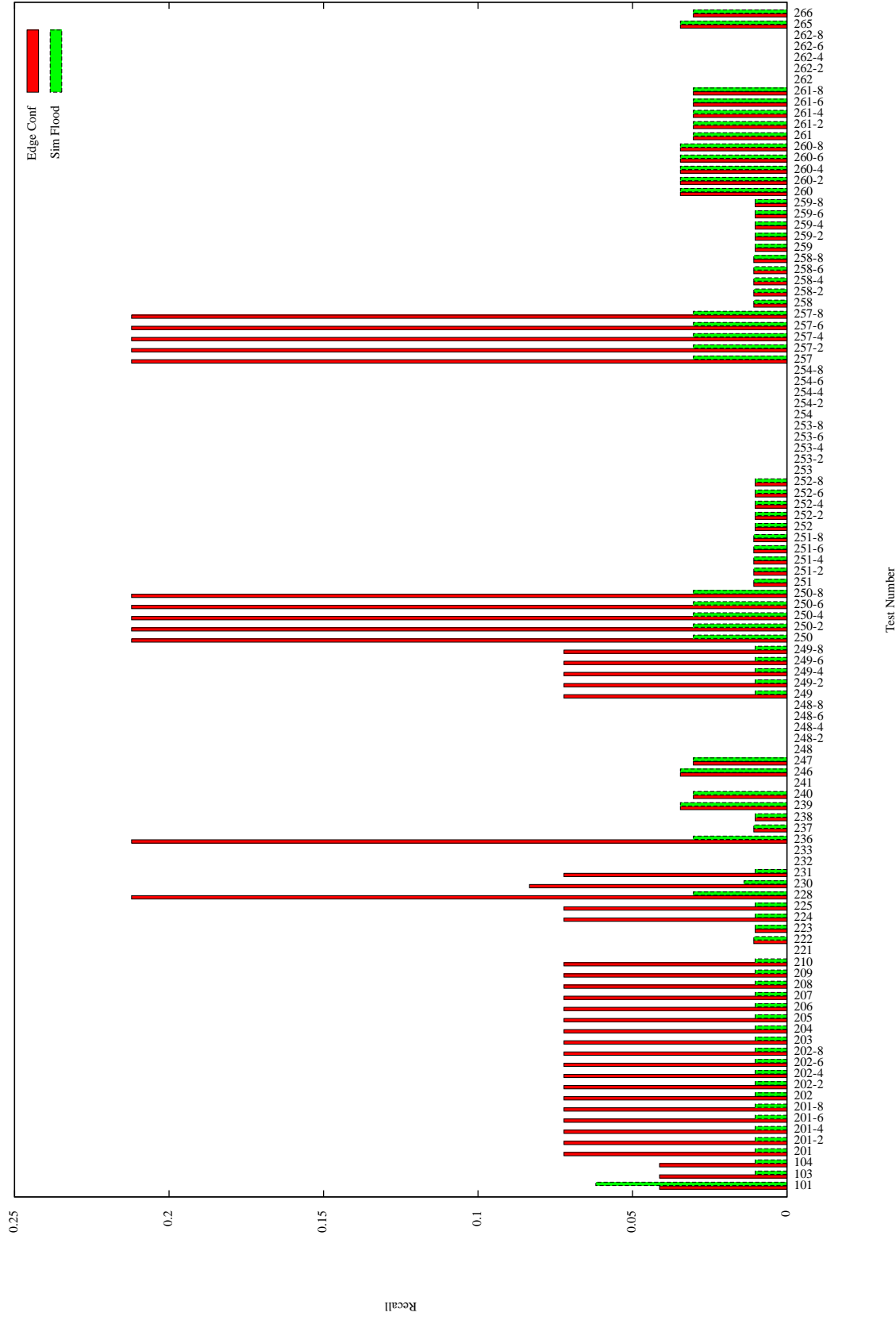}}
\caption{Comparison of Recall}
\label{fig:recall}
\end{sidewaysfigure}

By the same reasoning, Edge Flooding typically outperforms Similarity Flooding on the basis of recall. 
Because Edge Flooding return more results, it has more correct results, which gives a higher recall score. 
This is evident in Figure~\ref{fig:recall}.

\begin{sidewaysfigure}
\centering
\tiny{\input{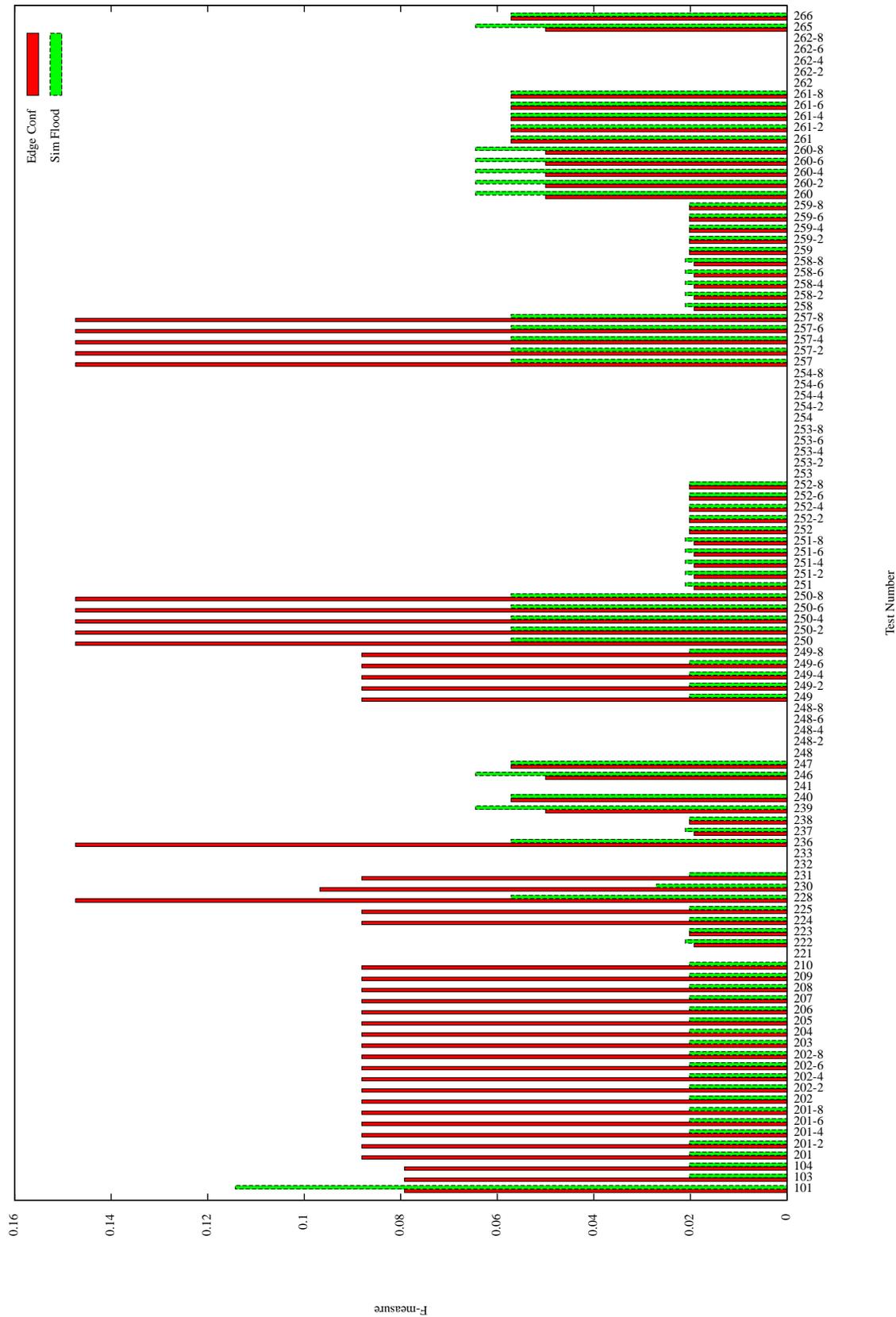}}
\caption{Comparison of F-measure}
\label{fig:fmeasure}
\end{sidewaysfigure}

The real question is: does the more aggressive approach pay off in the F-measure score, or does it extract too much of a penalty because of its lack of precision?
Figure~\ref{fig:fmeasure} shows that the Edge Flooding technique performs very well against Similarity Flooding, performing orders of magnitude better on many of  the tests.

\section{Conclusions \& Future Work}
\label{sec:conclusions}
In this paper, we have shown an improvement to the popular
Similarity Flooding technique for producing ontology alignment.
Our evaluation on the SEALS platform has shown promising results. While the overall
scores for both algorithms are very low, it should be noted that these algorithms
are not intended for stand-alone use, but rather as building blocks in more
sophisticated suites of ontology alignment tools. There is also more work
to be done in the future.
One avenue for improvement is to adjust the definition of edge confidence so that it 
includes information about the structural relationships between different object properties.
For example, object properties have parent-child relationships similar to ontological terms.
A new edge confidence function could take advantage of these relationships in order to 
consider a more structural approach to the alignment problem.


\section*{Acknowledgments}

We would like to acknowledge Dr. John Miller at the University of Georgia for providing us the opportunity to pursue this study through his Advanced Databases course. 

\bibliographystyle{abbrv}
\bibliography{refs} 

\begin{thebibliography}{10}

\bibitem{bcnn:2010:simulation}
J.~Banks, J.~S. Carson, B.~L. Nelson, and D.~M. Nicol.
\newblock {\em {Discrete-Event System Simulation}}.
\newblock Prentice Hall, 5th edition, 2010.

\bibitem{benzecri:1973:lsi}
J.-P. Benzécri.
\newblock {\em {L'Analyse des Donnees}}, volume~II.
\newblock Dunod, 1973.

\bibitem{cotterell:2012:oei}
M.~Cotterell, J.~Zheng, Q.~Sun, Z.~Wu, C.~Champlin, and A.~Beach.
\newblock {Facilitating Knowledge Sharing and Analysis in Energy Informatics
  with the Ontology for Energy Investigations (OEI)}.
\newblock In {\em Proceedings of the 2012 Conference on Energy Informatics},
  October 2012.

\bibitem{esteban:2010:executing}
M.~Esteban-Guti{\'e}rrez, R.~Garc{\'\i}a-Castro, and A.~G{\'o}mez-P{\'e}rez.
\newblock {Executing Evaluations over Semantic Technologies using the SEALS
  Platform}.
\newblock 2010.

\bibitem{euzenat:2007:ontology}
J.~Euzenat and P.~Shvaiko.
\newblock {\em {Ontology Matching}}, volume~18.
\newblock Springer Berlin, 2007.

\bibitem{hayes:2000:graph}
B.~Hayes.
\newblock {Graph Theory in Practice: Part I}.
\newblock {\em {American Scientist}}, 88(1):9--13, 2000.

\bibitem{kuhn:1955:hungarian}
H.~Kuhn.
\newblock {The Hungarian Method for the Assignment Problem}.
\newblock {\em {Naval Research Logistics Quarterly}}, 2(1-2):83--97, 1955.

\bibitem{lin:1998:information}
D.~Lin.
\newblock {An Information-Theoretic Definition of Similarity}.
\newblock In {\em Proceedings of the 15th International Conference on Machine
  Learning, 1998.}, volume~1, pages 296--304, 1998.

\bibitem{melnik:2002:similarity}
S.~Melnik, H.~Garcia-Molina, and E.~Rahm.
\newblock {Similarity Flooding: A Versatile Graph Matching Algorithm and its
  Application to Schema Matching}.
\newblock In {\em Proceedings of the 18th International Conference on Data
  Engineering}, pages 117--128. IEEE, 2002.

\bibitem{miller:1995:wordnet}
G.~Miller et~al.
\newblock {WordNet: A Lexical Database for English}.
\newblock {\em {Communications of the ACM}}, 38(11):39--41, 1995.

\bibitem{miller:2010:scalation}
J.~A. Miller, J.~Han, and M.~Hybinette.
\newblock {Using Domain Specific Language For Modeling and Simulation:
  {ScalaTion} as a Case Study}.
\newblock In {\em Proceedings of the 2010 Winter Simulation Conference}, pages
  741--752. ACM, December 2010.

\bibitem{mitzenmacher:2005:probability}
M.~Mitzenmacher and E.~Upfal.
\newblock {\em {Probability and Computing: Randomized Algorithms and
  Probabilistic Analysis}}.
\newblock Cambridge University Press, 2005.

\bibitem{odersky:2011:spec}
M.~Odersky.
\newblock {The Scala Language Specification, Version 2.9}.
\newblock Technical report, {Programming Methods Laboratory, EPFL}, May 2011.

\bibitem{rijsbergen:1979:ir}
C.~J.~V. Rijsbergen.
\newblock {\em {Information Retrieval}}.
\newblock {Butterworth-Heinemann}, Newton, MA, USA, 2nd edition, 1979.

\bibitem{winkler:1999:state}
W.~Winkler.
\newblock {The State of Record Linkage and Current Research Problems}.
\newblock {\em {Statistics of Income Division, In- ternal Revenue Service
  Publication R99/04}}, 1999.
\newblock Available from http://www.census.gov/srd/www/byname.html.

\bibitem{wrigley:2010:evaluating}
S.~Wrigley, K.~Elbedweihy, D.~Reinhard, A.~Bernstein, and F.~Ciravegna.
\newblock {Evaluating semantic search tools using the SEALS Platform}.
\newblock In {\em Proceedings of the International Workshop on Evaluation of
  Semantic Technologies (IWEST 2010), International Semantic Web Conference
  (ISWC2010), International Semantic Web Conference (ISWC2010), China}, 2010.

\bibitem{yujian:2007:levenshtein}
L.~Yujian and L.~Bo.
\newblock {A Normalized Levenshtein Distance Metric}.
\newblock {\em IEEE Transactions on Pattern Analysis and Machine Intelligence},
  29(6):1091--1095, June 2007.

\end{thebibliography}

\section*{Author Biographies} 
\vspace{8 pt}
\noindent \textbf{MICHAEL E. COTTERELL} is a Ph.D. student in Computer Science at the University of Georgia. 
He received his B.S. in Computer Science from UGA in May 2011. 
As an undergraduate, he served as the Vice President of the UGA Chapter of the Association for Computer Machinery (ACM) for two years. 
He also interned at the National Renewable Energy Lab, where he conducted research on formal ontologies for Energy Systems Integration and the Energy Informatics domain. 
His research interests include Simulation, Optimization, Semantic Web and Domain-Specific Languages with inter-disciplinary applications related to both Bioinformatics and Energy Informatics.
His email address is \href{mailto:mepcott@uga.edu}{mepcott@uga.edu}.
His curriculum vitae and list of publications are available at \url{http://michaelcotterell.com/}.

\vspace{8 pt}
\noindent \textbf{TERRANCE MEDINA} is a Master's student in Computer Science at the University of Georgia.  
His research interests include Agent-based Modeling and Simulation, Computational Music and Robot Control Architectures.
His email address is \href{mailto:medinat@uga.edu}{medinat@uga.edu}.

\onecolumn

\end{document}